# Belle II Executive Summary

on behalf of the U.S. Belle II Collaboration

D.M. Asner[1], H. Atmacan[2], Sw. Banerjee[3], J.V. Bennett[4], M. Bertemes[1], M. Bessner[5], D. Biswas[3], G. Bonvicini[6], N. Brenny[7], R.A. Briere[8], T.E. Browder[5], C. Chen[7], S. Choudhury[7], D. Cinabro[6], J. Cochran[7], L.M. Cremaldi[4], A. Di Canto[1], S. Dubey[5], K. Flood[5], B. G. Fulsom[9], V. Gaur[10], R. Godang[11], T. Gu[12], Y, Guan[2], J. Guilliams[4], C. Hadjivasiliou[9], O. Hartbrich[5], W.W. Jacobs[13], D.E. Jaffe[1], S. Kang[7], L. Kapitánová[12], C. Ketter[5], A. Khatri[7], K. Kinoshita[2], S. Kohani[5], H. Korandla[5], I. Koseoglu Sari[5], R. Kroeger[4], J. Kumar[8], K.J. Kumara[4], T. Lam[10], P.J. Laycock[1], L. Li[2], D. Liventsev[6], F. Meier[14], S. Mitra[7], A. Natochii[5], N. Nellikunnummel[1], K.A. Nishimura[5], E.R. Oxford[8], A. Panta[4], K. Parham[14], T.K. Pedlar[15], R. Peschke[5], L.E. Piilonen[10], S. Pokharel[4], S. Prell[7], H. Purwar[5], D.E. Ricalde Herrmann[6], C. Rosenfeld[16], D. Sahoo[7], D.A. Sanders[4], A. Sangal[2], V. Savinov[12], S. Schneider[14], J. Schueler[5], A.J. Schwartz[2], V. Shebalin[5], A. Sibidanov[5], Z.S. Stottler[10], J. Strube[9], S. Tripathi[5], S.E. Vahsen[5], G.S. Varner[5], A. Vossen[14], D. Wang[17], E. Wang[12], L. Wood[9], J. Yelton[17], Y. Zhai[7], and B. Zhang[5]

[1] Brookhaven National Laboratory, Upton, New York 11973
[2] University of Cincinnati, Cincinnati, Ohio 45221
[3] University of Louisville, Louisville, Kentucky 40292
[4] University of Mississippi, University, Mississippi 38677
[5] University of Hawaii, Honolulu, Hawaii 96822
[6] Wayne State University, Detroit, Michigan 48202
[7] Iowa State University, Ames, Iowa 50011
[8] Carnegie Mellon University, Pittsburgh, Pennsylvania 15215
[9] Pacific Northwest National Laboratory, Richland, Washington 99352
[10] Virginia Polytechnic Institute and State University, Blacksburg, Virginia 24061
[11] University of South Alabama, Mobile, Alabama 36688
[12] University of Pittsburgh, Pittsburgh, Pennsylvania 16260
[13] Indiana University, Bloomington, Indiana 47408
[14] Duke University, Durham, North Carolina 27708
[15] Luther College, Decorah, Iowa 52101
[16] University of South Carolina, Columbia, South Carolina 29208
[17] University of Florida, Gainesville, Florida 32611

## Introduction

**Belle II is the next-generation *B* Factory – a Super *B* Factory.** A *B*-factory experiment searches for physics beyond the Standard Model (SM) or "New Physics" at the "Rare and Precision Frontier." The experiment reconstructs large samples of $B$ mesons, charm hadrons, and tau leptons, and searches for the production of new particles (e.g., dark sector particles) in $e^+e^-$ collisions on and near the $\Upsilon$(4S) resonance. Historically, the flavor physics measurements from these experiments have played crucial roles in advancing our knowledge. For example, the previous-generation *B* factories, Belle and BaBar, *discovered CP violation in the $B$ system,* which was recognized by the 2008 Physics Nobel Prize. Belle II will make high-precision measurements of suppressed processes. If New Physics (NP) is present, its first signs may appear there. Precise measurements of suppressed decays can provide sensitivity to NP occurring at energy scales far above those that can be accessed in direct searches at high-$p_T$ hadron collider experiments. Mapping out and understanding non-SM physics will require the full range of Rare and Precision Frontier and Energy Frontier experiments. The complementarity between them, and the potential for uncovering NP they collectively provide, are great strengths of the US HEP program. Given the absence of non-SM discoveries at the LHC, the role of flavor physics experiments at the Rare and Precision Frontier has become even more crucial.

The Belle II experiment is a significant improvement over Belle and BaBar, which together have published almost 1200 papers. Belle II is expected to record 50 ab$^{-1}$ of data, which is *two orders of magnitude* greater than that recorded by BaBar and fifty times that of Belle. This large increase in luminosity is due to the use of low-emittance beams and strong vertical focusing, resulting in beam heights of only 60 nm - so-called "nano-beams." The new "Super-KEKB" accelerator is running with nano-beams and has achieved an instantaneous luminosity about 3.5 times greater than that achieved by the PEP-II accelerator at SLAC - and with a factor of five lower product of beam currents. Numerous world records for instantaneous luminosity, and daily and weekly integrated luminosities, have been set by this new, innovative machine. The SuperKEKB team with accelerator scientists from the US, EU and Asian laboratories are collaborating to optimize the path for accessing SuperKEKB's full potential.

**The Belle II detector is state-of-the-art, incorporating advanced detector technologies.** For example, it uses silicon pixels for the two innermost layers of the vertex detector. This improves the track impact parameter and vertex resolutions by a factor of two over those achieved at Belle and BaBar. [An early Belle II paper, using only 78 fb$^{-1}$ of data, measured the $D^+$ and $D^0$ lifetimes with the world's best precision.] In addition, the Belle II detector includes a new large-volume central tracker; a powerful particle identification detector (iTOP) based on Cherenkov light radiated in optically polished quartz bars; a new $K_L$ and muon detector (KLM); and state-of-the-art readout, trigger and DAQ systems. The US Belle II groups played a leading role in designing, constructing, and now operating the iTOP and KLM detectors. These detectors, and all other Belle II detectors, are collecting data and performing close to design specifications.

An $e^+e^-$ Super *B* factory has several advantages as compared to, e.g., a fixed target or hadron collider experiment. Background levels are generally low due to the small multiplicity of final state particles and the absence of pile-up; trigger efficiencies are near 100%; reconstruction efficiencies are essentially flat across Dalitz plots and as a function of decay time, and reconstruction of neutrals (i.e. photon, $\pi^0$, $K_S$, $K_L$; and subsequently $\eta$, $\rho^\pm$, $\omega$, $\eta'$, $a_0$, etc.) is relatively straightforward. Because the initial state is known and the detector is nearly hermetic, Belle II can reconstruct fully-inclusive final states and search for new particles via the “missing mass” technique. These searches can be performed irrespective of the lifetime of the particle or the final state into which it decays, even including “invisible” final states. A wide range of measurements in the *B* meson system using flavor-tagged, time-dependent observables can be carried out at threshold (on the ϒ(4S) resonance) where boosted $B\bar{B}$ pairs are produced in a quantum coherent state.

**Physics reach**

**The goal of Belle II is to uncover new physics beyond the Standard Model.** Belle II will pursue NP in many ways, for example: improving the precision of weak interaction parameters, particularly Cabibbo-Kobayashi-Maskawa (CKM) matrix elements and phases, and thus more rigorously test the CKM paradigm, measuring lepton-flavor-violating parameters, and performing unique searches for missing-mass dark matter events and fundamental tests of quantum coherence. As examples, the NP flavor structure with enhanced τ lepton couplings (e.g. $B \to D^{(*)} \tau \nu$) can be tested at higher precision than at a hadron collider, and the most sensitive methods to determine the CKM angle $\alpha/\varphi_2$ require measurements of asymmetries in modes such as $B \to \rho^+\rho^-$, $B \to \rho^\pm\pi^\mp$, and $B \to \pi^0\pi^0$. Theory sum rules (e.g., in $B \to K\pi$ decays) can be fully tested, as all terms - including those for decays having $\pi^0$s in the final state - can be measured. Belle II will access pure electroweak penguin processes such as $B \to K\nu\bar{\nu}$ whose rates and CP asymmetries are especially sensitive to NP. Belle II’s fully-inclusive measurements will have significantly reduced theory uncertainties compared to those made with exclusive final states. Belle II’s high trigger efficiency even for events with significant missing energy allows for effective searches for dark sector particles. Belle II’s measurement of the cross section for $e^+e^- \to \pi^+\pi^-$ will more precisely determine the leading-order hadronic contribution to the muon g−2 anomaly. Expected sensitivities to τ decays will be world-leading due to the production of clean $\tau^+\tau^-$ pairs at Belle II with minimal combinatorial and machine backgrounds. Many of these measurements can *only* be performed at Belle II, e.g., inclusive decays and absolute branching fractions are probably impractical to measure at a hadron collider. In all areas, Belle II will push far beyond the first-generation measurements of Belle and BaBar.

From 2015-2018, a series of workshops were held with experimentalists and theorists collaborating together to identify the most promising physics measurements for Belle II, culminating in *The Belle II Physics Book*[1]. These measurements address many fundamental questions of high energy physics. We list here some of the topics and corresponding measurements that Belle II is uniquely positioned to address (see the Snowmass Belle II Physics White Paper).

- **Testing violations of lepton flavor conservation and universality and understanding their origins**
  - The ratios of branching fractions $R(D^{(*)}) = B[B \to D^{(*)} \tau\nu]/B[B \to D^{(*)} \mu\nu]$ have shown some of the most significant discrepancies between SM predictions and measurements. The combined discrepancy is currently at the level of 3.4σ and indicates potential NP in leptonic couplings. Belle II will measure these ratios about three times more precisely than the current world averages. We can also probe inclusive semi-tauonic *B* decays, which have different theoretical uncertainties. The angular distributions of $B \to D^* l\, \nu$ (*l* =

[1] Belle II Physics Book, https://arxiv.org/abs/1808.10567, published as Prog Theor Exp Phys (2019), Issue 12

*e, μ, τ)* are sensitive to NP (e.g., the difference in forward-backward asymmetries $A_{FB}(B \to D^*\mu\nu) - A_{FB}(B \to D^*e\nu)$ ) and will be well measured by Belle II.

   - Belle II has good sensitivity to NP in decays involving internal loops (so-called "penguin" decays) such as $b \to s\,\nu\bar{\nu}$, $b \to s\,l^+l^-$ $(l = e, \mu, \tau)$, and $b \to s\gamma$; the current measured value for the ratio $R_K = B(B \to K\mu^+\mu^-) / B(B \to Ke^+e^-)$ differs from the SM by 3σ. We expect Belle II to discover $B^+ \to K^+\nu\bar{\nu}$, and to measure its branching fraction with about 10% uncertainty. The angular distributions of $B \to K^*\, l^+l^-$ (for all lepton flavors $l$ including $\tau$) will also be measured by Belle II.
   - Belle II will investigate possible lepton flavor violation (LFV) with τ decays to many final states including $\tau \to \mu\gamma$. Belle II will improve the sensitivity of this final state by a factor of six, while the sensitivity of dozens of modes will be improved by up to two orders of magnitude. *The observation of a single LFV tau decay mode would immediately establish physics beyond the SM.* Belle II can measure the electric dipole moment of the $\tau$ lepton with significantly greater precision than the current world-average; this measurement is sensitive to non-SM scalar couplings.

- **Checking the unitarity of the CKM matrix to high precision**
   - Belle II can measure the CKM angles $\beta/\varphi_1$, $\alpha/\varphi_2$, and $\gamma/\varphi_3$ with high precision. The least known of the three angles, $\alpha/\varphi_2$, can be determined with $B$ decays to $\rho\rho$, $\pi\pi$, and $\rho\pi$ final states. From the measurement of branching fractions and CP asymmetries of these decays (including those with neutral pions), Belle II can measure $\alpha/\varphi_2$ with a world-leading precision of less than 1°.
   - There is a long-standing discrepancy between the value of $|V_{cb}|$ measured in inclusive decays and that measured using exclusive decays. This 3.3σ discrepancy could indicate the presence of non-SM partial widths. A similar discrepancy exists between inclusively and exclusively measured values of $|V_{ub}|$. Belle II is in a unique position to understand and resolve these discrepancies, which are severely limiting precision measurements, as inclusive decays can only be studied in the experimentally clean $e^+e^-$ environment.
   - $|V_{us}|$ measurements from both kaon and τ decays are systematically smaller than the CKM unitarity constraint by ~3σ (the "Cabibbo angle anomaly"). Inclusive τ decays at Belle II provide an alternative way to measure $|V_{us}|$ with different systematic uncertainties than those of semileptonic kaon decay measurements.

- **Identifying new weak (CP-violating) phases in the quark sector**
   - CP asymmetries in decays proceeding via the penguin loop transitions $b \to s$ and $b \to d$ have high sensitivity to new weak phases from non-SM physics. Such asymmetries will be measured at Belle II in a variety of charged and neutral final states. Examples are the time-dependent CP asymmetries in $B^0 \to \eta' K^0_S$ and $B^0 \to \phi K^0_S$, which Belle II will measure with unique precision.
   - Measurements of the branching fraction ratio of charged and neutral $B \to K\pi$ decays deviate significantly from SM expectations (the "$K\,\pi$ puzzle"). Whether this discrepancy arises from additional SM contributions or from NP can be probed with an isospin sum rule involving $B \to K\pi$ branching fractions and CP asymmetries. Belle II can determine all terms with high precision, including the difficult-to-measure CP asymmetries in $B^0 \to K^0_S\pi^0$ that dominate the sensitivity of the sum rule (in fact, probably *only* Belle II can measure these asymmetries). As $B \to K\pi$ decays proceed via loop diagrams they can help identify non-SM contributions to electroweak penguin amplitudes.

  - Belle II will measure time-dependent CP violation in exclusive $b \to s\gamma$ decays (e.g., in $B \to K_S^0\pi^0\gamma$) that can arise from right-handed currents. We will also measure triple-product CP asymmetries in $B \to VV$ decays. Belle II will search for CP violation in many charm hadron decays, e.g., in $D^+\to \pi^+\pi^0$, where a nonzero CP asymmetry would unambiguously indicate NP.

- **Probing the existence of dark-sector particles at MeV to GeV mass scales**
  - A number of theoretical models predict a rich structure of dark-matter particles, axion-like-particles, and new gauge bosons in the MeV to GeV energy range. Such particles can be directly produced in $e^+e^-$ collisions at Belle II, but the experimental signatures can be challenging to identify, e.g., $e^+e^- \to \gamma$ + *nothing.* Belle II has implemented new trigger algorithms (such as a single photon trigger) to capture these elusive events. With such triggers, and using the missing-mass technique, Belle II has a unique ability to uncover these dark-sector particles.

- **Reducing the uncertainty in the theory prediction for the muon g−2 anomaly**
  - An important measurement of the US HEP program is that of the gyromagnetic ratio g of the muon, typically parameterized as the "anomaly" $a_\mu = (g-2)/2$. The current experimental value differs from the SM prediction by 4.2σ. The dominant theoretical uncertainty can be reduced by a more precise measurement of the $e^+e^- \to \pi^+\pi^-$ cross-section. The increased statistics at Belle II allows to reduce the systematic uncertainty in this measurement.

- **Understanding the role of QCD in the production and binding of new hadronic states of matter**
  - Exotic QCD states such as tetraquarks and QCD molecules can be produced at Belle II in several ways: near resonance by tuning the machine energy; through initial state radiation; or in $B$ decays such as $B \to X(3872)\ K$ and $B \to Z(4430)^+\ K$. With the ability to reconstruct essentially all neutral and charged particles, Belle II will play a unique role in the search for exotic states; especially for $b\bar{b}$ states that are currently accessible only at SuperKEKB and will be in the foreseeable future.

- **Unique studies in nuclear physics**
  - Measurements of di-hadron spin-momentum correlations at $B$ factories, such as the discovery of the Collins effect, have provided crucial input to the imaging of the partonic structure of nucleons. Precision data from Belle II will enable the extension of this physics program to multi-dimensional correlations of spin and momenta. Such measurements will provide important input to the design of a future electron-ion collider.

## US Role

**The US groups play a major role in the Belle II experiment.** Specifically, they played a *leading* role in the design, construction, and commissioning of the iTOP and KLM detectors, and they currently play a large role in their operations. They are leading critical machine-detector interface studies related to beam backgrounds, and the US hosts a Tier 1 GRID computing facility at BNL. US members have played leadership roles in Belle II management, serving as Belle II Spokesperson, Chair of the Institutional Board, Chair of the Speakers Committee, Chair of the Publication Committee, Software Coordinator, Data Production Manager, Deputy Run Managers, Physics Group Conveners, etc. US members had significant editorial responsibilities for *The Belle II Physics Book.* US postdocs and students are now analyzing the current dataset to produce limits on dark matter and rare *B* and *D* decays, improved measurements of charm baryon and meson lifetimes, and measurements of charm mixing and tau decays. US collaborators are heavily involved in detector and DAQ upgrades that are planned over the next ten years to further improve detector performance and data-taking efficiency.

Relative to U.S. investments in other forefront HEP experiments such as CMS, ATLAS, NovA, and DUNE, the US investment in Belle II is very modest. The overall financial commitment corresponds to only a *fraction of a percent* of the annual US HEP budget. Much of the cost of the detector, and all construction costs for the accelerator, were borne by the KEK laboratory. Given the NP discovery potential of Belle II, and the track record of the preceding Belle experiment, **the US investment in Belle II represents a very high value.**

## Timeline

Belle II began its initial physics run in 2019. Since then, SuperKEKB has carried out significant accelerator development, continuously improved its performance, and delivered 300 $fb^{-1}$ of integrated luminosity.

**Belle II's goal is to integrate 50 $ab^{-1}$ of data until 2035**[2]**.** This schedule includes two long shutdowns (LS1 and LS2) for detector upgrades. During LS1, Belle II will complete the installation of the second layer of the pixel-based inner vertex detector. For LS2, major upgrades to the detector are being considered, along with extensive upgrades to the accelerator. The detector upgrades being considered include: (a) replacing the MCP-PMTs in the iTOP detector with SiPMs to better cope with expected increases in luminosity; (b) completing the scintillator upgrade of the barrel KLM, to improve its hermiticity and timing; and (c) upgrading the iTOP and KLM front end electronics. Additional improvements being considered include CMOS vertexing and new electronics for the CDC. All planned detector upgrades are discussed in an accompanying Snowmass white paper. Similarly, planned upgrades of the SuperKEKB accelerator are described in a separate white paper.

We are exploring an upgrade to the accelerator that would result in polarization of the electron beam. A polarized beam would open up a new program of high-precision electroweak measurements. For example, the neutral current couplings of the *b* quark, *c* quark and muon can be measured several times more precisely than the current world averages. This upgrade could potentially be implemented before reaching the full 50 $ab^{-1}$ of data. Running with a polarized beam could allow Belle II to measure the anomalous magnetic moment of the τ lepton, which would be of great interest if the inconsistency between experiment and theory observed for the muon's magnetic moment persists. A Snowmass white paper describing the accelerator upgrade to facilitate a polarized beam is being submitted.

For the long-term future, if non-SM physics is observed in *B*, *D*, or τ decays, or a dark sector particle is discovered, then Belle II will continue its physics program to explore such phenomena. Studies are beginning on technical advances in accelerator R&D and the machine detector interface. These are synergistic with the FCC-ee project and could lead to instantaneous luminosities as high as 2 ×

[2] https://www-superkekb.kek.jp/Luminosity_projection.html

$10^{36}/cm^2/sec$. Such an increase in luminosity would require a major R&D program for the detector, in order to improve its radiation hardness and its ability to reject backgrounds.[3]

## Summary

**The physics program of Belle II has outstanding potential for discovering non-SM physics over the next decade.** We strongly agree with the assessment of the European Strategy Group[4] that *"The quest for dark matter and the exploration of flavor and fundamental symmetries are crucial components in the search for new physics."* In addition, if unambiguous non-SM physics signals are observed in flavor-physics measurements made at other experiments, independent measurements at Belle II will be important. The broad program of fundamental weak interaction measurements, the current hints of non-SM flavor signals, as well as the exciting possibility of New Physics discoveries in searches unique to Belle II warrant enhanced U.S. investment in the Belle II and SuperKEKB and their upgrades.

[3] KEK Roadmap 2021, pg 25, https://www.kek.jp/wp-content/uploads/2021/06/KEKroadmap2021_E.pdf
[4] European Strategy Group, Physics Briefing Book, arxiv:1910:11775